# Overlapping-gate architecture for silicon Hall bar MOSFET devices in the low electron density and high magnetic field regime

Laurens H. Willems van Beveren[1, a], Kuan Y. Tan[2], Nai-Shyan Lai[2], Oleh Klochan[3], Andrew S. Dzurak[2], and Alex R. Hamilton[3]

[1]Centre for Quantum Computation and Communication Technology, School of Physics, The University of Melbourne, VIC 3010, Australia

[2]Centre for Quantum Computation and Communication Technology, School of Electrical Engineering & Telecommunications, The University of New South Wales, Sydney 2052, Australia

[3]School of Physics, The University of New South Wales, Sydney 2052, Australia

[a]laurensw@unimelb.edu.au

**Abstract.** A common issue in low temperature measurements of enhancement-mode metal-oxide-semiconductor (MOS) field-effect transistors (FETs) in the low electron density regime is the high contact resistance dominating the device impedance. In that case a voltage bias applied across the source and drain contact of a Hall bar MOSFET will mostly fall across the contacts (and not across the channel) and therefore magneto-transport measurements become challenging. However, from a physical point of view, the study of MOSFET nanostructures in the low electron density regime is very interesting (impurity limited mobility [1], carrier interactions [2,3] and spin-dependent transport [4]) and it is therefore important to come up with solutions [5,6] that work around the problem of a high contact resistance in such devices (c.f. Fig. 1 (a)).

## Introduction

In this work, the authors report the fabrication and study of silicon-based Hall bar MOSFET devices in which an overlapping-gate architecture, as shown in Fig. 1 (b), allows four-terminal measurements of low electron density 2D systems, while maintaining a high electron density at the ohmic contacts, using the lead gate electrodes (grey) in Fig. 1 (c) at both source and drain contact [7]. The Hall bar device shown in Fig. 1 (b) was fabricated by electron-beam lithography (EBL) and local oxidation of the bottom layer of aluminum (green) to form the $Al_xO_y$ dielectric that electrically insulates the top layer of aluminum (grey).

## High magnetic field sweeps

To explore spin-related physics in these EBL patterned Hall bar MOS devices the carriers, in this case electrons, need to have definite spin states, with an energy splitting larger than the thermal energy and the intrinsic linewidth of each Landau level. For example, for spin-selective tunneling of quantum-Hall edge channels at a tunnel barrier, spin-resolved Landau levels are required [8].

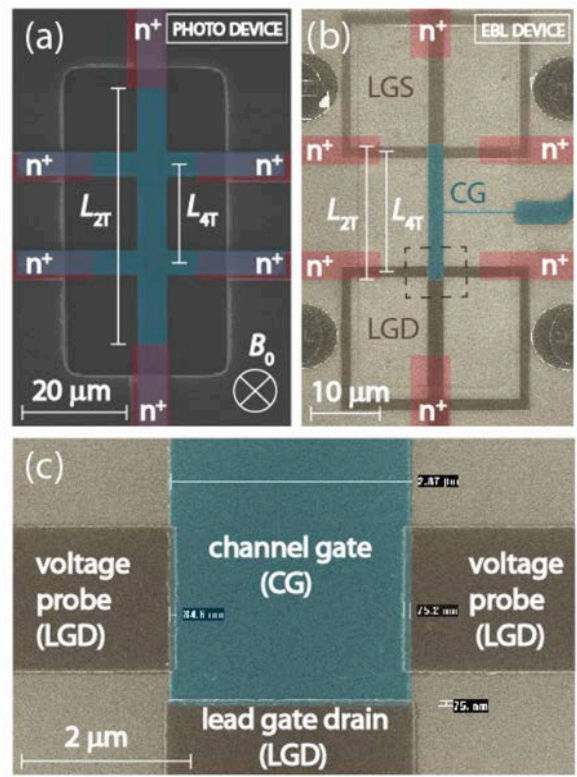

Fig. 1: (a) SEM image of a conventional, single-gated, Hall bar MOSFET and (b) the Hall bar with overlapping-gate architecture. (c) Zoom-in of the drain region.

Furthermore, in the case of silicon, the lifting of the spin-valley degeneracy could play an important role in the electronic transport properties in such devices and acts as an additional degree of freedom. In order to explore whether this spin-resolved Landau level regime can be achieved in silicon we reduce the 2DEG electron density $n$ as well as the Landau level filling factor value $v$ by increasing the magnetic field strength to values as high as possible. Our previous (unpublished) studies have only been carried out to magnetic field strengths of 8 T and yielded Landau level filling factor values of $v = 4$ for the lowest density possible.

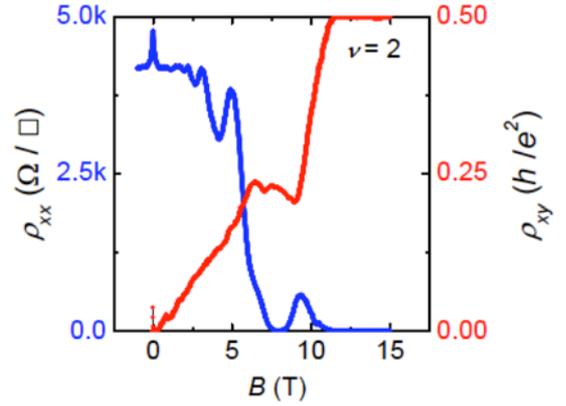

Here, we have performed magnetic field sweeps up to 15 T at fixed gate voltage (or constant 2DEG density $n$) and hope to observe the Zeeman splitting of each individual Landau level. The resulting data is shown in Fig. 2 for both $\rho_{xx}$ and $\rho_{xy}$, respectively. For these measurements all gate electrodes were biased simultaneously to avoid gate-to-gate leakage.

This data demonstrates the oscillatory behavior in the longitudinal resistivity $\rho_{xx}$ and the plateaus-like behavior in the transverse (Hall) resistance $\rho_{xy}$ driven by the formation of Landau levels with increasing magnetic field strength. The gate voltage applied for these field sweeps was $V_G = 1$V, which corresponds to a 2DEG density $n = 0.8 \times 10^{12}$ /cm$^2$. This density was derived from the linear part of $\rho_{xy}$ versus $B$ in the low field regime.

Fig. 2: (a) Magnetic field sweeps of $\varrho_{xx}$ and $\varrho_{xy}$ up to $B = 15$ T for a gate voltage $V_G = 1$ V, corresponding to a 2DEG density $n = 0.8 \times 10^{12}$ /cm$^2$. The $v = 2$ plateau corresponds to $\varrho_{xy} = h/2e^2$.

Around $B = 0$ T a clear weak-localization peak can be seen in $\rho_{xx}$ and only for magnetic fields bigger than 8 T the minimum values of $\rho_{xx}$ drop down to zero resistance. Associated with this minimum is the step-like structure in $\rho_{xy}$ which marks the $v = 4$ regime. For fields higher than 12 T there is a clear quantized Hall plateau at $v = 2$ ($\rho_{xy} = h/2e^2 = 12.9$ k$\Omega$).

### Landau level fan diagram

To map out the full parameter space of magnetic field and 2DEG density we have measured the Landau level fan diagram, which is shown in Fig. 3. This diagram depicts $\rho_{xx}$ as a function of both $B$ and $V_G$ over the full range of these parameters, i.e. $B = 0$ to 15 T and $V_G = 0.8$ to 2.0 V. This data shows that filling factor values of $v = 2$ can be achieved already for magnetic field values of 8 T, provided the electron density (gate voltage) is low enough. Furthermore, it seems the lowest Landau level broadens with increasing magnetic field, which could indicate the Zeeman splitting becoming larger than the width of the Landau level, the latter determined by the mobility.

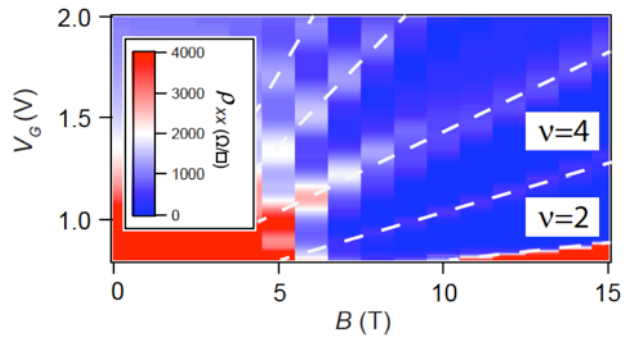

Fig. 3: Landau level fan diagram for a magnetic field ranging from $B = 0$ to 15 T and a gate voltage ranging from $V_G = 0.8$ to 2.0 V. Dashed white lines are only a guide to the eye.

## Energy scales

The Zeeman energy at 15 T corresponds to $E_Z$ = 1.73 meV (assuming a g-factor of 2.0 for the 2DEG and ignoring the oscillatory enhancement of the g-factor [9]), which exceeds the thermal energy of $k_BT$ = 7.6 μeV at 100 mK by many orders. The spacing between the Landau levels at 15 T is given by $\hbar\omega_c$ = 17.15 meV, which is the dominant energy scale. The peak mobility of the EBL device was measured to be $\mu$ = 2500 cm$^2$/Vs, which results in a collision time for the electrons of $\tau_e = \mu m_e^*/e$ = 0.26 ps ($m_e^*$ = 0.19$m_e$ in silicon). The associated mean free path $l_e$ is then $l_e = v_e\tau_e$ = 26 nm, assuming a velocity of $v_e = \hbar k_F/m_e^*$ = 0.96 x 10$^5$ m/s using $k_F^2 = 4\pi n/(g_s g_v)$ where $g_s = g_v$ = 2, respectively. The collision time of the electron can be converted into a Landau level broadening of width $\Delta E \sim \hbar/2\tau_e$ = 1.22 meV. Therefore, it is likely that the increase in width of the lowest Landau level is due to the Zeeman splitting as this energy starts to exceed the level broadening for magnetic fields larger than 2 T, which seems consistent with our data.

## Summary

Comparison with devices made using a standard single gate (c.f. Fig. 1 (a)) show that measurements can be performed at much lower densities and higher channel resistances, despite a reduced peak mobility [7]. We also observe a voltage threshold shift which is attributed to negative oxide charge, injected during the electron-beam lithography processing. Hall bar data obtained at magnetic fields of 15 T show that Landau level filling factors of $\upsilon$ = 2 can readily be achieved in the low-density regime. Here, either the spin or valley degeneracy is lifted. Based on the above calculated energy scales we think that the $\upsilon$ = 2 plateau is due to two unresolved valley states and not due to unresolved spin states. This is consistent with previously reported spin-resolved Landau levels [10].

## Acknowledgements

We acknowledge support from the Australian Research Council Discovery scheme (grant DP0772946), the ARC Centre of Excellence for Quantum Computation and Communication Technology (project number CE110001027), and the US National Security Agency and the US Army Research Office under contract number W911NF-08-1-0527. A.R.H. acknowledges an ARC Professorial Fellowship.